\documentclass[ reprint, amsmath,amssymb,aps,]{revtex4-1}
\usepackage{graphicx}
\usepackage{dcolumn}
\usepackage{bm}
\usepackage{pst-all}
\usepackage{float}
\usepackage{lineno}
\usepackage{pst-all}

 \usepackage{ifthen}
 \usepackage{times}

\begin{document}

\preprint{APS/123-QED}

\title{Baryonic resonances close to the \bf{$\mathrm{\bar{K}N}$} threshold: the case of \bf{$\Lambda(1405)$} in pp collisions}
\author{
G.~Agakishiev$^{6}$, A.~Balanda$^{3}$, D.~Belver$^{18}$, A.~Belyaev$^{6}$, 
J.C.~Berger-Chen$^{8}$, A.~Blanco$^{2}$, M.~B\"{o}hmer$^{9}$, J.~L.~Boyard$^{16}$, P.~Cabanelas$^{18}$, 
S.~Chernenko$^{6}$, A.~Dybczak$^{3}$, E.~Epple$^{8}$, L.~Fabbietti$^{8,\ast}$, 
P.~Finocchiaro$^{1}$, P.~Fonte$^{2,b}$, J.~Friese$^{9}$, I.~Fr\"{o}hlich$^{7}$, T.~Galatyuk$^{7,c}$, 
J.~A.~Garz\'{o}n$^{18}$, R.~Gernh\"{a}user$^{9}$, K.~G\"{o}bel$^{7}$, M.~Golubeva$^{12}$, D.~Gonz\'{a}lez-D\'{\i}az$^{d}$, 
F.~Guber$^{12}$, M.~Gumberidze$^{16}$, T.~Heinz$^{4}$, T.~Hennino$^{16}$, R.~Holzmann$^{4}$, 
A.~Ierusalimov$^{6}$, I.~Iori$^{11,f}$, A.~Ivashkin$^{12}$, M.~Jurkovic$^{9}$, B.~K\"{a}mpfer$^{5,e}$, 
T.~Karavicheva$^{12}$, I.~Koenig$^{4}$, W.~Koenig$^{4}$, B.~W.~Kolb$^{4}$, G.~Kornakov$^{18}$, 
R.~Kotte$^{5}$, A.~Kr\'{a}sa$^{17}$, F.~Krizek$^{17}$, R.~Kr\"{u}cken$^{9}$, H.~Kuc$^{3,16}$, 
W.~K\"{u}hn$^{10}$, A.~Kugler$^{17}$, A.~Kurepin$^{12}$, V.~Ladygin$^{6}$, R.~Lalik$^{8}$, 
S.~Lang$^{4}$, K.~Lapidus$^{8}$, A.~Lebedev$^{13}$, T.~Liu$^{16}$, L.~Lopes$^{2}$, 
M.~Lorenz$^{7}$, L.~Maier$^{9}$, A.~Mangiarotti$^{2}$, J.~Markert$^{7}$, V.~Metag$^{10}$, 
B.~Michalska$^{3}$, J.~Michel$^{7}$, C.~M\"{u}ntz$^{7}$, R.~M\"unzer$^8$, L.~Naumann$^{5}$, Y.~C.~Pachmayer$^{7}$, 
M.~Palka$^{3}$, Y.~Parpottas$^{15,14}$, V.~Pechenov$^{4}$, O.~Pechenova$^{7}$, J.~Pietraszko$^{7}$, 
W.~Przygoda$^{3}$, B.~Ramstein$^{16}$, A.~Reshetin$^{12}$, A.~Rustamov$^{7}$, A.~Sadovsky$^{12}$, 
P.~Salabura$^{3}$, A.~Schmah$^{a}$, E.~Schwab$^{4}$, J.~Siebenson$^{8,\ast}$, 
S.~Spataro$^{g}$, B.~Spruck$^{10}$, H.~Str\"{o}bele$^{7}$, J.~Stroth$^{7,4}$, C.~Sturm$^{4}$, 
A.~Tarantola$^{7}$, K.~Teilab$^{7}$, P.~Tlusty$^{17}$, M.~Traxler$^{4}$, R.~Trebacz$^{3}$, 
H.~Tsertos$^{15}$, T.~~Vasiliev$^{6}$, V.~Wagner$^{17}$, M.~Weber$^{9}$, C.~Wendisch$^{5}$, 
J.~W\"{u}stenfeld$^{5}$, S.~Yurevich$^{4}$, Y.~Zanevsky$^{6}$}

\affiliation{
(HADES collaboration) \\\mbox{$^{1}$Istituto Nazionale di Fisica Nucleare - Laboratori Nazionali del Sud, 95125~Catania, Italy}\\
\mbox{$^{2}$LIP-Laborat\'{o}rio de Instrumenta\c{c}\~{a}o e F\'{\i}sica Experimental de Part\'{\i}culas , 3004-516~Coimbra, Portugal}\\
\mbox{$^{3}$Smoluchowski Institute of Physics, Jagiellonian University of Cracow, 30-059~Krak\'{o}w, Poland}\\
\mbox{$^{4}$GSI Helmholtzzentrum f\"{u}r Schwerionenforschung GmbH, 64291~Darmstadt, Germany}\\
\mbox{$^{5}$Institut f\"{u}r Strahlenphysik, Helmholtz-Zentrum Dresden-Rossendorf, 01314~Dresden, Germany}\\
\mbox{$^{6}$Joint Institute of Nuclear Research, 141980~Dubna, Russia}\\
\mbox{$^{7}$Institut f\"{u}r Kernphysik, Goethe-Universit\"{a}t, 60438 ~Frankfurt, Germany}\\
\mbox{$^{8}$Excellence Cluster 'Origin and Structure of the Universe' , 85748~Garching, Germany}\\
\mbox{$^{9}$Physik Department E12, Technische Universit\"{a}t M\"{u}nchen, 85748~Garching, Germany}\\
\mbox{$^{10}$II.Physikalisches Institut, Justus Liebig Universit\"{a}t Giessen, 35392~Giessen, Germany}\\
\mbox{$^{11}$Istituto Nazionale di Fisica Nucleare, Sezione di Milano, 20133~Milano, Italy}\\
\mbox{$^{12}$Institute for Nuclear Research, Russian Academy of Science, 117312~Moscow, Russia}\\
\mbox{$^{13}$Institute of Theoretical and Experimental Physics, 117218~Moscow, Russia}\\
\mbox{$^{14}$Frederick University, 1036~Nicosia, Cyprus}\\
\mbox{$^{15}$Department of Physics, University of Cyprus, 1678~Nicosia, Cyprus}\\
\mbox{$^{16}$Institut de Physique Nucl\'{e}aire (UMR 8608), CNRS/IN2P3 - Universit\'{e} Paris Sud, F-91406~Orsay Cedex, France}\\
\mbox{$^{17}$Nuclear Physics Institute, Academy of Sciences of Czech Republic, 25068~Rez, Czech Republic}\\
\mbox{$^{18}$LabCAF. Dpto. F\'{\i}sica de Part\'{\i}culas, Univ. de Santiago de Compostela, 15706~Santiago de Compostela, Spain}\\ 
\\
\mbox{$^{a}$ also at Lawrence Berkeley National Laboratory, ~Berkeley, USA}\\
\mbox{$^{b}$ also at ISEC Coimbra, ~Coimbra, Portugal}\\
\mbox{$^{c}$ also at ExtreMe Matter Institute EMMI, 64291~Darmstadt, Germany}\\
\mbox{$^{d}$ also at Technische Universit\"{a}t Darmstadt, 64289~Darmstadt, Germany}\\
\mbox{$^{e}$ also at Technische Universit\"{a}t Dresden, 01062~Dresden, Germany}\\
\mbox{$^{f}$ also at Dipartimento di Fisica, Universit\`{a} di Milano, 20133~Milano, Italy}\\
\mbox{$^{g}$ also at Dipartimento di Fisica Generale and INFN, Universit\`{a} di Torino, 10125~Torino, Italy}\\
\\
\mbox{$^{\ast}$ corresponding authors: laura.fabbietti@ph.tum.de, johannes.siebenson@ph.tum.de}
} 






\begin{abstract}
We present an analysis of the $\Lambda(1405)$ resonance produced in the reaction $p+p\rightarrow \Sigma^{\pm}+\pi^{\mp}+K^++p $ at $3.5\, \mathrm{GeV}$ kinetic beam energy measured with HADES at GSI.
The two charged decay channels $\Lambda(1405)\rightarrow\Sigma^{\pm}\pi^{\mp}$ have been reconstructed for the first time in p+p collisions. The efficiency and acceptance-corrected spectral shapes show a peak position clearly below $1400\, \mathrm{MeV/c}^2$. We find a total production cross section of $\sigma_{\Lambda(1405)}=9.2 \pm 0.9 \pm 0.7 ^{+3.3}_{-1.0}\,\mu$b. The analysis of its polar angle distribution suggests that the $\Lambda(1405)$ is produced isotropically in the p-p center of mass system.
\end{abstract}

\pacs{14.20.Jn, 13.30.-a}
\maketitle


\section{Introduction}
Lying slightly below the $\mathrm{\bar{K}-N}$ threshold ($\approx30\,\mathrm{MeV/c^2}$), the broad $\Lambda(1405)$ resonance is considered to be strongly linked to the antikaon-nucleon interaction. Hence the understanding of this resonance is mandatory to address the issue of the interaction. The $\Lambda(1405)$ was first observed experimentally by studying its presence in the $\Sigma\pi$ exit channel in $\mathrm{K^-}$-induced reactions \cite{Als61}. From a theoretical point of view the $\Lambda(1405)$ is treated within a coupled channel approach, based on chiral dynamics, in which the low-energy $\mathrm{\bar{K}-N}$ interactions can be handled \cite{theoSumm}. In this Ansatz the $\Lambda(1405)$ appears naturally as a dynamically generated resonance, resulting from the superposition of two components: a quasi-bound $\mathrm{\bar{K}-N}$ state and a $\Sigma\pi$ resonance.\\
At present, the molecule-like character of the $\Lambda(1405)$ is commonly accepted. However, the contribution of the $\Sigma\pi$ channel to the formation process is still discussed controversially. 
Indeed phenomenological approaches different from chiral-SU(3) predictions \cite{Esma} support the hypothesis that the $\Lambda(1405)$ can be considered as a pure $\mathrm{K^-p}$ quasi-bound state and suggest experimental methods to test this Ansatz.
In general, models can be constrained above the  $\mathrm{\bar{K}-N}$ threshold by $\mathrm{K^-p}$ scattering data 
 and by the measurements of the $\mathrm{Kp}$, $\mathrm{Kn}$ scattering lengths extracted from kaonic atoms as shown in \cite{sid:2011,weis:2011}. 
Below threshold, the only experimental observable related to the $\mathrm{\bar{K}-N}$ interaction is the $\Lambda(1405)$ spectral shape extracted from the decays $\Lambda(1405)\xrightarrow{\approx100 \%} (\pi\Sigma)^0$. 
The authors of reference  \cite{Nach99}  predict for the $\Lambda(1405)$ in the reaction $\mathrm{\gamma+p \rightarrow \Lambda(1405)+K^0}$ that the spectral functions of the three final states $\Sigma^{-} \pi^{+}/\Sigma^0\pi^0/\Sigma^+\pi^-$ should differ
because of the interference of the isospin 0 and 1 channels.
 In fact, the measured invariant 
mass distributions of  the $\Sigma\pi$ states have different shapes \cite{Mor10}, which also vary as a function of the photon energy,
but the observed shifts of the distributions are not compatible with the theoretical predictions.

Furthermore, the approach \cite{Maga05} predicts that  the coupling of the $\Lambda(1405)$ resonance to the quasi-bound $\mathrm{\bar{K}-N}$ state and
the $\Sigma\pi$ pole depends  on the initial state configuration. The observed line shape and 
pole position of the $\Lambda(1405)$ is expected to vary for different reactions. Data exploiting pion 
\cite{Thom73} and kaon \cite{Hem85} beams are scarce, and the reaction $\mathrm{p+p \rightarrow p+ \Lambda(1405)(\rightarrow \Sigma^0+ \pi^0}) + K^+ $ 
 has been studied hitherto only by the ANKE experiment \cite{L1405Anke} at a beam momentum of $3.65\, \mathrm{GeV/c}$. 
 
Based on the analysis of the reaction  $\mathrm{p+p \rightarrow p+ K^+ +  (\Sigma +\pi)^0}$ at 3.5 GeV kinetic beam energy, measured 
by HADES \cite{hadesSpectro}, we present the first data on the decay of the $\Lambda(1405)$ resonance into the $\Sigma^{\pm}\pi^{\mp}$
 final states. The spectral shapes, the polar production angle, and the production 
cross-section of the $\Lambda(1405)$ are discussed.
 
\section{Analysis} 
\subsection{Signal extraction}
The properties of the $\Lambda(1405)$ resonance are studied in the associated production together with a proton and a $K^+$ followed by the decay into $\Sigma^{\pm}+\pi^{\mp}$ pairs, where a branching ratio of $33.3\,\%$ for each decay channel is assumed: 
\begin{flalign} 
\begin{pspicture}(1,-0.3)
 \psline[ArrowInside=-]{->}(3.7,-0.4)(3.7,-0.75)(4.2,-0.75)
     \psline[ArrowInside=-]{->}(4.5,-1.0)(4.5,-1.3)(4.9,-1.3)          
\end{pspicture}
 p+p  \xrightarrow{3.5GeV}\Lambda(1405)& + K^{+}+p   \label{L1405Production}\\
                                        &\Sigma^{\pm}+\pi^{\mp}               \nonumber  \\   
                                        &\qquad \pi^{\pm}+n.   \nonumber   
\end{flalign}                                              
\begin{figure}[]
	\centering
		\includegraphics[width=0.47\textwidth]{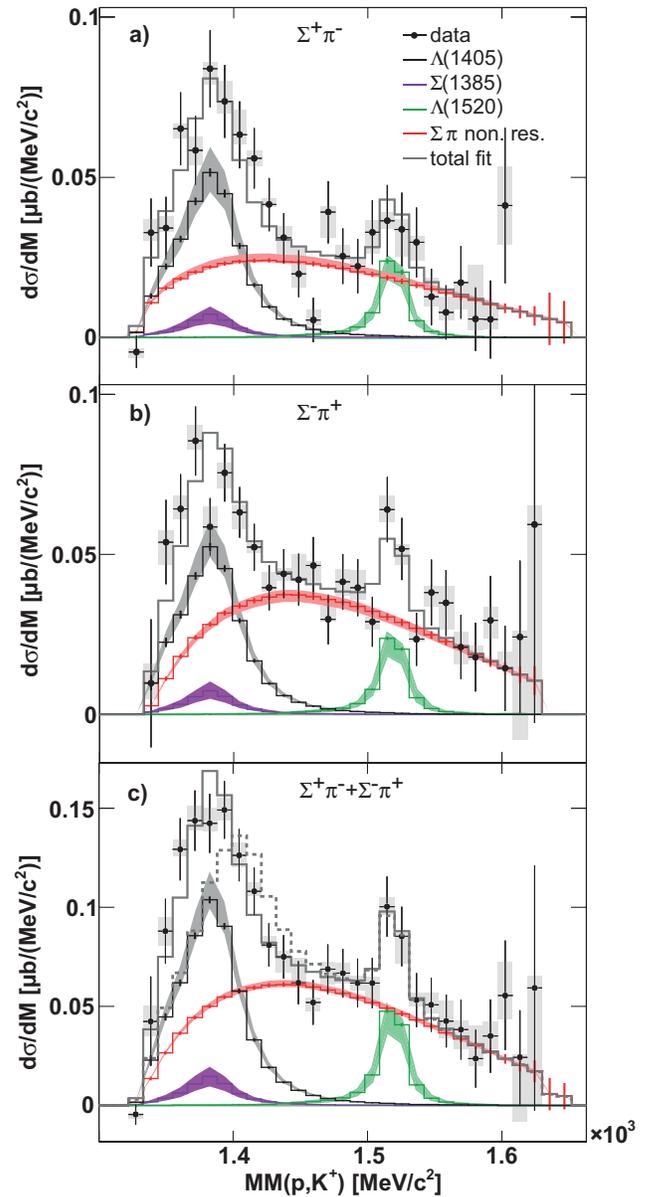}
	\caption{(Color online) Missing mass $\mathrm{MM(p,K^+)}$ distributions for events attributed to the $\Sigma^+\pi^-$ decay channel a) and to the $\Sigma^-\pi^+$ decay channel b). Panel c): sum of both spectra from panels a) and b). The gray dashed histogram shows the sum of all simulated channels if the $\Lambda(1405)$ is simulated with its nominal mass of $\mathrm{1405\, MeV/c}^2$. Colored histograms in the three panels indicate the contributions of the channels (\ref{L1405Production}-\ref{SMnonres_channel}) obtained from simulations. Data and simulations are acceptance and efficiency corrected. The gray boxes indicate systematic errors.}
	\label{fig:Corr}
\end{figure} 
The assumption about the branching ratios of the $\Lambda(1405)$ decays is motivated by the consideration of isospin symmetries \cite{ose99} and does not take into account the interference between the two isospins states $1$ and $0$.
For an exclusive analysis, all charged particles ($\mathrm{p,\,K^+,\,\pi^+,\,\pi^-}$) in the final state have been identified employing the momentum dependent $\mathrm{dE/dx}$ and velocity information \cite{S1385:2011}. The neutron appearing in the reaction (\ref{L1405Production}) has been reconstructed via the missing mass to the four charged particles $p,~\pi^{\pm},~\pi^{\mp},~K^+$ and
has been selected via a $2.4\,\sigma$ cut around the nominal neutron mass (see Fig.1 in \cite{Valencia2010}). The intermediate $\Sigma^+$ and $\Sigma^-$ hyperons have been reconstructed via the missing mass to the proton, the $\mathrm{K^+}$ and either the $\pi^-$ or the $\pi^+$ (see Fig.~4 in \cite{Valencia2010}). $3\,\sigma$ mass cuts around the nominal masses of the $\Sigma^+$ and $\Sigma^-$ hyperons allow to extract the $\Lambda(1405)$  signal corresponding to the two decay modes into $\Sigma^+\pi^-$ and $\Sigma^-\pi^+$. After the subtraction of the misidentification background due to the limited kaon identification \cite{S1385:2011}, the  $\Lambda(1405)$  spectral shape for both decay channels can be analyzed in the missing mass spectra to the proton and the $\mathrm{K^+}$, $\mathrm{MM(p,K^+)}$. 
Figure~\ref{fig:Corr} shows the  $\mathrm{MM(p,K^+)}$ distributions for the $\Sigma^+\pi^-$ a) and $\Sigma^-\pi^+$ b) decay channels. The black dots correspond to the experimental data.
Together with the reaction~(1) the following contributions have been considered:\\

$p+p\rightarrow$
\begin{align}
&\Sigma(1385)^0+p+K^+\stackrel{ 12 \% }{\rightarrow}\Sigma^{\pm}+\pi^{\mp}+p+K^+ \label{S13850_channel},\\
&\Lambda(1520)+p+K^+ \stackrel{28 \%}{\rightarrow}\Sigma^{\pm}+\pi^{\mp}+p+K^+\label{L1520_channel},\\
&\Sigma^{+}+\pi^{-}+p+K^+ \label{SPnonres_channel}, \\
&\Delta^{++}(1232)+\Sigma^-+K^+\stackrel{100 \%}{\rightarrow}p+\pi^++\Sigma^-\label{SMnonres_channel}+K^+.
\end{align}
Full scale simulations of these channels have been carried out and the relative contribution of each of them has been evaluated from a simultaneous fit to the two missing mass distributions $\mathrm{MM(p,K^+,\pi^{\pm})}$ together with the two $\mathrm{MM(p,K^+)}$ distributions \cite{Valencia2010}.
The area around $1400\,\mathrm{ MeV/c}^2$ for $\mathrm{MM(p,K^+)}$ has been excluded from the fit in order to not bias the finally extracted shape of the $\Lambda(1405)$ resonance by the simulated $\Lambda(1405)$ shape. In total, a normalized $\chi^2$ value of $\mathrm{\chi^2/ndf\approx\,1.3}$ has been obtained. 
Figure~\ref{fig:Corr} shows the contributions of the different channels together with their incoherent sum (gray histogram, solid line).

\begin{figure}[t]
	\centering
		\includegraphics[width=0.46\textwidth]{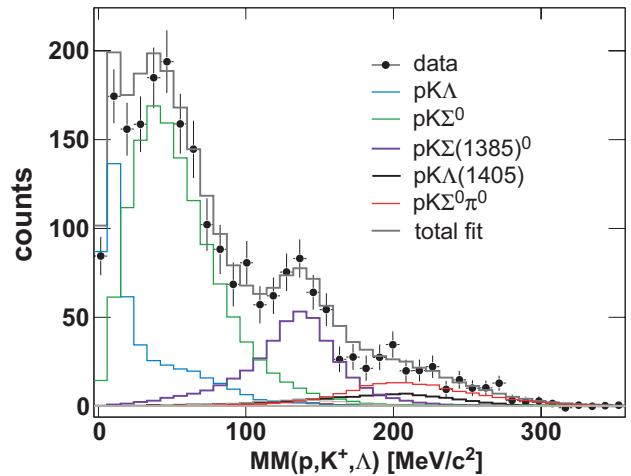}
	\caption{(Color online) Missing mass to the  p, $\Lambda$  and $\mathrm{K^+}$ particles. The colored histograms show the contributions of the simulated channels 6-9. The experimental data are uncorrected. See the text for details.}
	\label{fig:MMall}
\end{figure}

The data and the full-scale simulations shown in Fig.~\ref{fig:Corr} a)-c) are corrected for acceptance and efficiency and the statistical and systematic errors for both, experimental data and simulations are displayed. 
The finite geometrical acceptance of HADES and the total reconstruction efficiency have been calculated using full-scale simulations of the channels (1-5), including the correct angular distribution for these channels as described  below.
The systematic errors shown in Fig.~\ref{fig:Corr} have been obtained by varying the selection cuts by $\pm\,20\%$ and the angular distribution of the simulated reactions (4) and (5) by $\pm\,30\%$. \\
The experimental data (black dots) in Fig.~\ref{fig:Corr} a) and b) show two distinct peak structures. The one slightly below $1400\,\mathrm{MeV/c^2}$ is mainly due to the $\Lambda(1405)$ resonance, whereas the second peak around $1500\,\mathrm{ MeV/c^2}$ is attributed to the $\Lambda(1520)$ resonance. 
The relative contribution of $\Lambda(1405)$ and $\Sigma(1385)^0$ can not be determined by fitting the simulations to the experimental data since their mass spectra overlap completely. 
However, the contribution of the $\Sigma(1385)^0$ can be inferred from the analysis of the 
$\Sigma(1385)^0\rightarrow \Lambda +\pi^0$  decay. 
Figure~\ref{fig:MMall} shows the $\mathrm{MM(p,K^+,\Lambda)}$ distribution where the $\Sigma(1385)^0$ contribution corresponds to a $\pi^0$ peak, while the $\Lambda(1405)$ corresponds to a broader distribution located at higher mass due to the additional $\gamma$ present in the decay $\Lambda(1405)\rightarrow\Sigma^0\pi^0$. 
 The experimental data presented here are not corrected for acceptance and efficiency. The histograms shown together with the experimental data correspond to full-scale simulations of relevant reaction channels with a hyperon in the final state which have been fitted to the data as described in \cite{EppleFabbietti:2011}; the results of the fit are presented.
The most relevant channels visible in Fig.~\ref{fig:MMall} are: \\
$p+p\rightarrow$
\begin{align}
&\Sigma^0+p+K^+{\rightarrow}\Lambda + \gamma+p+K^+ \label{1channel},\\
&\Sigma(1385)^0+p+K^+\stackrel{ 87 \% }{\rightarrow}\Lambda + \pi^0+p+K^+ \label{2_channel},\\
&\Lambda(1405)+p+K^+\stackrel{ 33.3 \% }{\rightarrow}\Lambda + \pi^0+\gamma+p+K^+ \label{3_channel},\\
&\Sigma^0+\pi^0+p+K^+{\rightarrow}\Lambda + \pi^0+\gamma+p+K^+ \label{4_channel}.
\end{align}
The missing mass distributions of the reactions (\ref{3_channel}) and (\ref{4_channel})  are almost identical and the statistics is limited in the missing mass range where their contributions is expected. This translates into a large uncertainty in the fitted result for the relative yield of these two channels and leads to a large error in the estimate of the production cross-section of the $\Lambda(1405)$. On the other hand, the signal corresponding to the $\Sigma(1385)^0$ is more significant, due to the large branching ratio of the decay (7).
The fit method discussed in \cite{EppleFabbietti:2011} allowed to extract a ratio for the $\Sigma(1385){^0}$ and $\Lambda(1405)$ cross-section, where the large error is mainly due to the inaccuracy of the determination of the $\Lambda(1405)$ contribution. 
A slightly extended version of this analysis was carried out, where as an additional systematic check the nominal mass of the $\Lambda(1405)$ was assumed to be $1385\,\mathrm{MeV/c^2}$. 
Instead of considering  the $\Sigma(1385){^0}$ to $\Lambda(1405)$ ratio, the cross-section for the ${\Sigma(1385)^0}$ was estimated from the fit of the experimental data via the different simulated channels and found to be $\sigma_{\Sigma(1385)^0}=\,5.56\pm0.48^{+1.94}_{-1.06}\,\mu\mathrm{b}$, while the results for the $\Lambda(1405)$ vary between $0$ and $14\,\mathrm{\mu b}$.
The contribution of the $\Sigma(1385)^0$ can be suppressed by selecting in the $\mathrm{MM(p,K^+,\Lambda)}$ distribution the region above $195\,\mathrm{ MeV/c^2}$ (see Fig.~\ref{fig:MMall}).  However,  only $100$ counts are left after this selection and this statistics contain the contribution by the channels (7-9). Hence, an analysis of the $\Lambda(1405)$ line shape exploiting the neutral $\Sigma^0\pi^0$ decay channel is not possible with this data sample.
The uncertainty of the angular distribution of $\Sigma(1385)^0$ has been included in the systematic error of the production cross-section by considering the two extreme cases of an isotropic distribution  for the $\Sigma(1385)^0$ and of the same angular distribution as measured for the $\Sigma(1385)^+$ \cite{S1385:2011}.
The obtained cross-section for the  $\Sigma(1385)^0$ has been utilized to calculate the relative contribution to the $\mathrm{MM(p,K^+)}$ spectrum shown in Fig.~\ref{fig:Corr} by the magenta histogram. 
The systematic uncertainties have been propagated accordingly. 
Note that the low contribution by the $\Sigma(1385)^0$ to the yield shown in Fig.~\ref{fig:Corr} is correlated with the small branching ratio of channel (2).

Figure~\ref{fig:Corr} c) shows the sum of the distributions from the two final states $\Lambda(1405)\rightarrow\Sigma^{\pm}\pi^{\mp}$ for experimental data and simulations.
The good agreement between the corrected experimental data and the simulation (gray histogram, solid line) is obtained by simulating the $\Lambda(1405)$ as a relativistic s-wave Breit-Wigner distribution with a width of $50\, \mathrm{MeV/c}^2$ and a pole mass of $1385\,\mathrm{ MeV/c}^2$.
 Using instead the nominal mass of $1405\,\mathrm{ MeV/c}^2$ results in the gray dashed histogram in Fig.~\ref{fig:Corr} c) which fails obviously to describe the experimental $\Lambda(1405)$ peak structure. The difference is expressed by the two $\chi^2$ values of $0.6$ and $2.1$ respectively.
 
A good fit is also obtained, if the cross-section of the $\Sigma(1385)^0$ is not fixed and the $\Lambda(1405)$ is generated with its PDG values for mass and width. In this scenario the production cross-sections obtained for the $\Lambda(1405)$ and $\Sigma(1385)^0$ are approximately $3$ and $50\,\mu\mathrm{b}$ respectively. The cross-section of the $\Sigma(1305)^0$ would then largely exceed the value of the cross-section extracted from the neutral decay analysis and also the measured cross-section for the $\Sigma(1385)^+$ ($22\, \mu\mathrm{b}$) in the same data sample.  This contradicts the findings at higher energies reported in \cite{Kle70}, where  cross-sections of $7\, \mu\mathrm{b}$ and $15\, \mu\mathrm{b}$ are measured for the $\Sigma(1385)^0$ and $\Sigma(1385)^+$ hyperons, respectively, produced in p+p collisions at $6\, \mathrm{GeV}$.
These arguments strongly disfavor this second scenario. 
 On the other hand, one should mention that, analog to the S\"oding mechanism \cite{So}, interferences of resonant and non-resonant amplitudes with the same exit channel can cause an apparent shift of the peak of the spectral distribution without a change of the $\Lambda(1405)$ pole mass. Our efficiency and acceptance-corrected experimental data are hence a perfect tool to test different theoretical models.
 

\subsection{Angular Distributions}
\label{angDist}

\begin{figure}[b]
	\centering
		\includegraphics[width=0.39\textwidth]{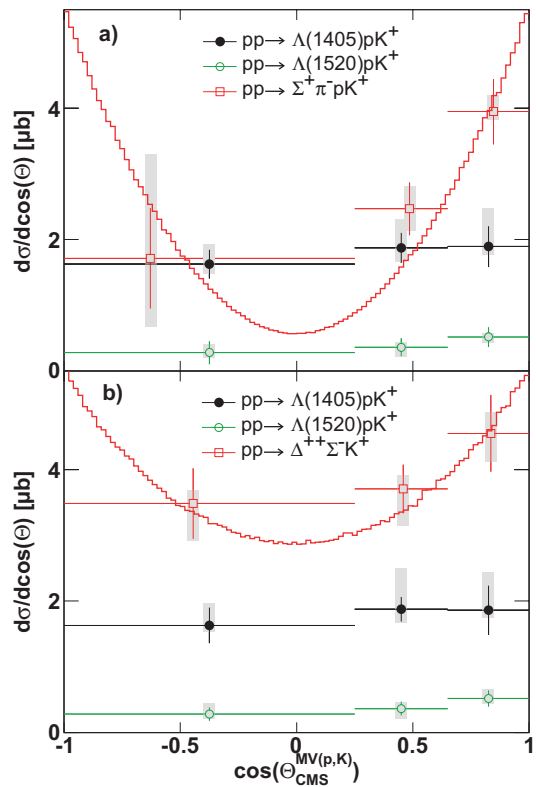}
	\caption{(Color online) Differential cross-section for the different simulated reaction channels as a function of $\mathrm{cos\left(\theta^{MV(p,K}_{CMS}\right)}$ for the $\Sigma^+\pi^-$ decay channel a) and the $\Sigma^-\pi^+$ decay channel b). The cross-sections are obtained by integrating the simulations, which have been fitted to the experimental data. The gray boxes indicate the systematic errors. The solid histograms show the angular distribution used to fold the simulation of the reactions $\mathrm{p+p \rightarrow \Sigma^+ + \pi^- +p +K^+}$ and $p+p \rightarrow \Delta^{++} + \Sigma^- +K^+$, respectively.}
	\label{fig:angular}
\end{figure}

The different sources contributing to the missing mass spectra shown in Fig.~\ref{fig:Corr}
have been studied in terms of their polar angle ($\theta$) in the p+p center of mass system. 
The results provide on the one hand constraints on possible production mechanisms of the $\Lambda(1405)$
and allow to compute the acceptance corrections. The polar angle of the missing momentum vector to the ($\mathrm{p-K}^+$) system, $\mathrm{MV(p,K)}$, has been investigated.
  The resulting angular distribution of $\mathrm{cos\left(\theta^{MV(p,K)}_{CMS}\right)}$
has been been divided into three intervals. Each of the three above resulting 
subsamples is treated in the same way as described for the angle integrated event sample,
meaning that the simulated distributions of MM(K,p) have been fitted to the experimental ones. 
In the fits the polar angle distribution of the $\Sigma(1385)^0$ was assumed to be the same as reported in \cite{S1385:2011}, for the $\Sigma(1385)^+$. Corrected experimental $\mathrm{MM(p,K^+)}$ distributions like those in Fig.~\ref{fig:Corr} 
have been obtained in each bin of $\mathrm{cos\left(\theta^{MV(p,K)}_{CMS}\right)}$. The cross-sections of the reactions 
\ref{L1405Production}-\ref{SMnonres_channel} are the integrals of
the simulated distributions. The results are plotted as a function of $\mathrm{cos\left(\theta^{MV(p,K)}_{CMS}\right)}$ in Fig.~\ref{fig:angular}. The shown systematic errors are obtained by varying the different selection cuts as 
described above by $\pm 20$ $\%$.
\\
The results suggest that the yield in the mass region of the $\Lambda(1405)$ as well as the $\Lambda(1520)$ is produced rather isotropically, whereas the production in the channels (\ref{SPnonres_channel}) and (\ref{SMnonres_channel}) is anisotropic. The observed angular distributions have been included by folding the simulations of the reactions (\ref{SPnonres_channel}) and (\ref{SMnonres_channel}) with the red curves shown in panel a) and b) of Fig.~\ref{fig:angular} that represent Legendre polynomials of second degree. The $\Lambda(1405)$ and $\Lambda(1520)$ production has been simulated isotropically. These simulations have been used to produce the acceptance and efficiency corrections applied to the data shown in Figs.~\ref{fig:Corr} and \ref{fig:angular}. The curvatures of the Legendre polynomials shown in panels a) and b) in Fig.~\ref{fig:angular} have been varied within $30\,\%$ and the simulations have been folded with the obtained angular distributions. The resulting uncertainty has been included in the systematic errors shown by the gray shaded boxes in Figs.~\ref{fig:Corr} and \ref{fig:angular}. \\
The simulation model obtained from this analysis has been tested for several other observables and overall a good agreement with the experimental data is obtained \cite{NPA2011}. \\
Finally, our corrected spectra allow to extract cross-sections for the channels (\ref{L1405Production}-\ref{SMnonres_channel}). This is again done by integrating the simulated spectra and using the statistical errors from the experimental data. We get the following values:
\begin{align}
&\sigma_{pp\rightarrow\Lambda(1405)pK^+}=9.2\pm0.9\pm0.7^{+3.3}_{-1.0} \mbox{ $\mu$b} \label{cs1}, \\
&\sigma_{pp\rightarrow\Lambda(1520)pK^+}=5.6\pm1.1\pm0.4^{+1.1}_{-1.6} \mbox{ $\mu$b} \label{cs2},  \\
&\sigma_{pp\rightarrow\Sigma^+\pi^-pK^+}=5.4\pm0.5\pm0.4^{+1.0}_{-2.1} \mbox{ $\mu$b}  \label{cs3}, \\
&\sigma_{pp\rightarrow\Delta^{++}\Sigma^-K^+}=7.7\pm0.9\pm0.5^{+0.3}_{-0.9} \mbox{ $\mu$b} \label{cs4}.
\end{align}  
\begin{figure}[t]
	\centering
		\includegraphics[width=0.46\textwidth]{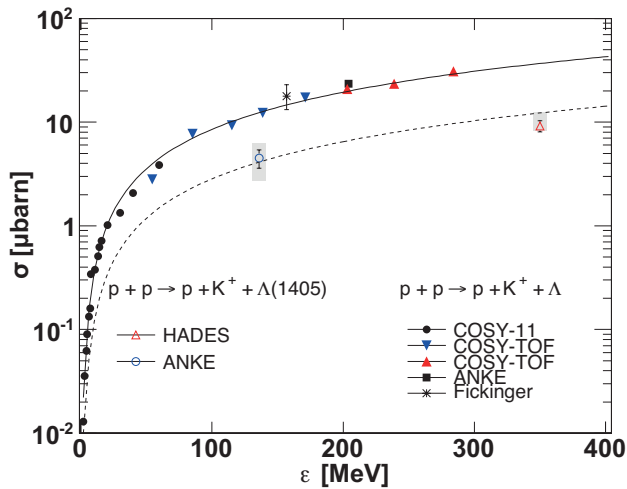}
	\caption{Compilation of production cross-sections as a function of the excess energy $\epsilon$ for the reactions $\mathrm{p+p\rightarrow \Lambda+K^++p}$ from \cite{Wiss} where the proper quotations of the references listed in the legend can be found, $\mathrm{p+p\rightarrow \Lambda(1405)+K^++p}$ from ANKE \cite{L1405Anke} and this work (data point at an excess energy $\epsilon=\,100 \,\mathrm{MeV}$). See the text for details.}
	\label{fig:cross}
\end{figure}
The first error gives the statistical error, the second one gives the systematic error from the normalization to the p+p elastic cross-section and the last error is obtained from the systematic variations mentioned above.
As demonstrated in \cite{NPA2011} the total yield of the non-resonant contribution shown in Fig.~\ref{fig:Corr} panel b) is attributed to the excitation of a $\Delta^{++}$ resonance, while for the charge conjugated final state, the expected contribution by $N^*$ resonances can not be verified probably due to their large width.

 Figure~\ref{fig:cross} shows a compilation of the production cross-section as a function of the excess energy $\epsilon$ for the channels $\mathrm{p+p\rightarrow \Lambda+K^++p}$ \cite{Wiss} and $\mathrm{p+p\rightarrow \Lambda(1405)+K^++p}$ \cite{L1405Anke}. The solid curve shown in Fig.~\ref{fig:cross} corresponds to the parametrization of the $\Lambda$ production discussed in \cite{Wiss}. The dashed curve has been obtained by scaling the $\Lambda$ parametrization by a factor $1/3$. One can see that the dependence of the $\Lambda(1405)$ production cross-section upon the excess energy seems to follow the same behaviour as exhibited by $\Lambda$ production in p+p collisions.

\section{Summary}
The $\Sigma^+\pi^-$ and $\Sigma^-\pi^+$ decay channels of the $\Lambda(1405)$ have been studied for the first time in p+p collisions at $\mathrm{3.5 \,GeV}$. These results can substantially contribute understanding the nature of $\Lambda(1405)$ which is considered as the key stone for the study of the $\mathrm{\bar{K}-N}$ interaction. The study of the spectral shape extracted from the decay into the $\Sigma^0\,+\,\pi^0$ channel was not possible, due to the limited statistics.
 The obtained results indicate a shift of the $\Lambda(1405)$ resonance in p+p reactions to values clearly below $1400\,\mathrm{MeV/c}^2$ with a maximum of the mass distribution around $1385~\mathrm{MeV/c}^2$ for both decay channels. If one considers the values of the masses of the two $\mathrm{\bar{K}-p}$  and  $\Sigma\pi$ poles recently constrained more precisely by new data on kaonic-hydrogen \cite{Ike12}, our result suggests that in p+p collisions the contribution of the $\Sigma\pi$ pole to the formation of the $\Lambda(1405)$ might be dominant. 
 
 The here presented mass distributions differ from the one measured in $\gamma$- and K-induced reactions \cite{Mor10,Hem85} and also from the measurement of the $\Lambda(1405)\rightarrow\Sigma^0\pi^0$ in p+p at 2.85GeV collisions \cite{L1405Anke} and the corresponding theoretical study \cite{Gen08}. The fact that the $\Sigma\pi$ spectra connected to the $\Lambda(1405)$ resonance strongly differ among different reactions could indicate that the production mechanism depends upon the entrance channel or that interference effects between the different channels contributing to the observed final state occur.  One way or the other precise measurements exploiting different beams together with a theory able to describe at the same time all the experimental results are necessary to clarify the situation. 

The angular distributions of the $\Lambda(1405)$ and $\Lambda(1520)$ in the CMS are rather isotropic, suggesting a large momentum transfer in the production mechanism. A total cross-section of $\sigma_{\Lambda(1405)}=\,9.2 \pm0.9\pm0.7^{+3.3}_{-1.0}\, \mu\mathrm{b}$ was reconstructed, which is about a factor of two smaller than the cross-section extracted for the $\Sigma(1385)^+$ \cite{S1385:2011}. The comparison of the $\Lambda(1405)$ production cross-section with the systematics measured as a function of the excess energy for the $\mathrm{\Lambda pK^+}$ final state shows that the two available data points are consistent with the phase-space trend measured for the $\Lambda$ production. A comparable production cross-section has been extracted for the reaction $pp\rightarrow\Delta^{++}\Sigma^-K^+$ underlining the role played by $\Delta$ resonances in the production mechanisms discussed here.
Considering the hypothesis that the $\Lambda(1405)$ might be a doorway for the formation of kaonic bound states, the results presented here are a necessary bench mark for theory to correctly address the formation of states like $\mathrm{ppK^-}$ produced in p+p collisions further decaying into $p+\Lambda$. \\

 The HADES collaboration gratefully acknowledges the support by the grants LIP Coimbra, Coimbra (Portugal) PTDC/FIS/113339/2009, SIP JUC Cracow, Cracow (Poland): N N202 286038 28-JAN-2010 NN202198639 01-OCT-2010, FZ Dresden-Rossendorf (FZD), Dresden (Germany) BMBF 06DR9059D, TU MŸnchen, Garching (Germany) MLL M\"unchen: DFG EClust 153, VH-NG-330 BMBF 06MT9156 TP5 GSI TMKrue 1012 NPI AS CR, Rez, Rez (Czech Republic) MSMT LC07050 GAASCR IAA100480803, USC - S. de Compostela, Santiago de Compostela (Spain) CPAN:CSD2007-00042, Goethe-University, Frankfurt (Germany): HA216/EMMI HIC for FAIR (LOEWE) BMBF:06FY9100I GSI F\&E.
 

\end{document}